\documentclass[conference]{IEEEtran}
\IEEEoverridecommandlockouts
\usepackage{cite}
\usepackage{amsmath,amssymb,amsfonts}
\usepackage{algorithmic}
\usepackage{graphicx}
\usepackage{textcomp}
\usepackage{xcolor}
\usepackage{algorithm}
\usepackage{algorithmic}
\usepackage{subfigure}
\usepackage{multirow}
\usepackage{booktabs}
\usepackage{hyperref}

\hypersetup{
	colorlinks=true,        
	linkcolor=blue,         
	citecolor=green,          
	urlcolor=magenta     
}

\def\BibTeX{{\rm B\kern-.05em{\sc i\kern-.025em b}\kern-.08em
		T\kern-.1667em\lower.7ex\hbox{E}\kern-.125emX}}
\begin{document}
	
	\title{Benchmarking Machine Learning Methods for Distributed Acoustic Sensing 
	}

	\author{ 
		\IEEEauthorblockN{Shuaikai Shi}
	\IEEEauthorblockA{\textit{School of Physics} \\
	\textit{Nanjing University}\\
	\textit{Nanjing,   China}\\
			huanianss@qq.com}
		\and
		\IEEEauthorblockN{Qijun Zong}
		\IEEEauthorblockA{\textit{School of Physics} \\
			\textit{Nanjing University}\\
			\textit{Nanjing,   China}\\
			qjzong2017@sinano.ac.cn\\
		corresponding author}
		
	}

	\maketitle
	
	\begin{abstract}
	Distributed acoustic sensing (DAS) technology represents an innovative fiber-optic-based sensing methodology that enables real-time acoustic signal monitoring through the detection of minute perturbations along optical fibers. This sensing approach offers compelling advantages, including extensive measurement ranges, exceptional spatial resolution, and an expansive dynamic measurement spectrum.
	The integration of machine learning (ML) paradigms presents transformative potential for DAS technology, encompassing critical domains such as data augmentation, sophisticated preprocessing techniques, and advanced acoustic event classification and recognition. By leveraging ML algorithms, DAS systems can transition from traditional data processing methodologies to more automated and intelligent analytical frameworks.
	The computational intelligence afforded by ML-enhanced DAS technologies facilitates unprecedented monitoring capabilities across diverse critical infrastructure sectors. Particularly noteworthy are the technology's applications in transportation infrastructure, energy management systems, and Natural disaster monitoring frameworks, where the precision of data acquisition and the reliability of intelligent decision-making mechanisms are paramount.
	This research critically examines the comparative performance characteristics of classical machine learning methodologies and state-of-the-art deep learning models in the context of DAS data recognition and interpretation, offering comprehensive insights into the evolving landscape of intelligent sensing technologies.
	\end{abstract}
	

	\section{Introduction}
	
In the forefront of contemporary sensing technologies, distributed acoustic sensing (DAS) \cite{dasreview} is spearheading a perception revolution by transforming fiber infrastructure into an intelligent, continuous acoustic sensing system that transcends the limitations of traditional point sensors. This innovative technology leverages unique optical physics principles to convert optical fibers into ultra-high-precision acoustic sensors, enabling real-time, continuous, and high-resolution monitoring of minute environmental acoustic disturbances, thereby providing unprecedented perceptive capabilities across diverse domains including infrastructure safety, geophysical exploration, and security monitoring. Unlike conventional discrete sensing methodologies, DAS technology constructs a continuous, dynamic, and highly sensitive environmental monitoring network by capturing subtle acoustic wave variations along optical fibers, achieving a remarkable degree of precision in acoustic signal acquisition and analysis that opens new technological frontiers for scientific research and engineering applications.

The contemporary technological landscape has been profoundly transformed by the exponential advancement of computational intelligence, with machine learning (ML) emerging as a pivotal paradigm that transcends traditional computational boundaries. Artificial intelligence has demonstrated unprecedented capabilities in computational domains, revolutionizing complex problem-solving methodologies across interdisciplinary research frontiers. Particularly in domains such as natural language processing and computer vision, sophisticated algorithmic approaches have catalyzed remarkable technological breakthroughs, fundamentally reimagining the potential of intelligent systems.
The convergence of machine learning with specialized sensing technologies represents a critical inflection point in technological innovation. By integrating intelligent computational frameworks with advanced sensing infrastructures like DAS, researchers can now develop sophisticated signal processing methodologies that dramatically enhance data interpretation capabilities. These intelligent algorithmic approaches enable unprecedented precision in acoustic signal analysis, transforming raw sensory data into actionable insights through advanced computational intelligence techniques.
This research critically examines the performance characteristics and computational efficacy of machine learning methodologies—ranging from classical statistical learning approaches to cutting-edge deep learning architectures—in the context of acoustic signal recognition and interpretation, offering a comprehensive analytical framework for understanding the evolutionary trajectory of intelligent sensing technologies.

In the realm of machine learning, feature extraction stands as a critical methodology for enhancing classification model performance. Traditional approaches have long utilized sophisticated statistical techniques such as time–frequency images, Mel cepstrum, and wavelet transforms to derive meaningful insights from complex datasets. Concurrently, the emergence of deep learning has catalyzed innovative mapping strategies that transform time series data into rich two-dimensional visual representations. Particularly noteworthy are techniques like short-time Fourier transform and Gramian Angular Field (GAF), which ingeniously convert one-dimensional temporal signals into feature-rich images that can be effectively processed by advanced neural network architectures. By bridging the gap between raw data and intelligent feature representation, these methodologies enable more nuanced, accurate, and powerful machine learning models capable of extracting profound patterns and relationships hidden within intricate datasets.

The remainder of this paper is organized as follows. Section \ref{sec:2} presents a comprehensive overview of the relevant machine learning methods, while Section \ref{sec:3} details the utilized datasets. The experimental results are systematically discussed in Section \ref{sec:4}, and Section \ref{sec:5} synthesizes the key conclusions drawn from our research.

\section{Related Machine Learning Methods}
	\label{sec:2}
In this section, we introduces several related methods, including conventional machine learning approaches and the advanced deep neural networks.

\subsection{Conventional Machine Learning Approaches}

Conventional machine learning approaches encompass a diverse array of powerful algorithms designed to address complex classification and prediction challenges. Logistic regression (LR) \cite{lr}, a foundational statistical method, models the probability of class membership through a sigmoid function, providing a straightforward yet robust approach to binary and multiclass classification problems. Support Vector Machines (SVM) \cite{svm} represent a more advanced technique, focusing on identifying optimal hyperplanes that maximize the margin between different class boundaries. By transforming data into higher-dimensional spaces and employing kernel methods, SVMs excel in handling non-linear relationships and achieving precise classification, particularly in scenarios with complex, intricate data distributions.

K-Nearest Neighbors (KNN) \cite{knn} and eXtreme Gradient Boosting(XGBoost) \cite{xgboost} offer complementary strategies for machine learning tasks. KNN, a non-parametric algorithm, operates on the principle of proximity-based classification, where new instances are categorized based on the majority vote or average of their k nearest neighbors in the feature space. This approach provides remarkable flexibility and intuitive interpretability, especially when dealing with datasets lacking clear parametric assumptions. XGBoost, an advanced ensemble learning technique, leverages gradient boosting to create powerful predictive models by sequentially combining multiple weak learners. Its sophisticated algorithm handles complex feature interactions, manages missing data efficiently, and incorporates regularization techniques, making it particularly effective in achieving high predictive accuracy across various domains, from financial forecasting to medical diagnostics.

\subsection{Advanced Deep Neural Networks}
Advanced Deep Neural Networks have revolutionized machine learning by introducing increasingly sophisticated architectures capable of handling complex pattern recognition and representation learning tasks. Multilayer perceptrons (MLP) serve as a foundational neural network architecture, consisting of multiple fully connected layers that enable non-linear transformations of input data. Convolutional neural networks (CNN)\cite{cnn} build upon this foundation, introducing specialized architectures for spatial data processing, particularly excelling in image recognition tasks through their unique convolutional and pooling layers. Residual networks (ResNet)\cite{resnet} further advanced this paradigm by introducing skip connections that allow deeper network architectures to be trained more effectively, overcoming the vanishing gradient problem and enabling the construction of extremely deep neural networks with unprecedented performance.

Recurrent Neural Networks, particularly long short-term memory (LSTM) \cite{lstm} networks, represent a critical advancement in handling sequential and time-series data, with their unique gating mechanisms enabling long-term dependency modeling. The Transformer \cite{transformer} architecture marked a pivotal moment in deep learning, introducing self-attention mechanisms that dramatically improved performance in natural language processing and sequence modeling tasks\cite{hsi}. Most recently, the Mamba \cite{mamba2} architecture has emerged as a promising alternative to Transformers, offering improved efficiency in processing long-sequence data through its state-space model approach. These advanced neural network architectures demonstrate the field's continuous evolution, progressively addressing computational challenges and expanding the boundaries of machine learning capabilities across diverse domains, from computer vision and natural language processing to complex predictive modeling and pattern recognition.

\subsection{Evaluation Metrics}
When evaluating the performance of the ML methods mentioned above, several classification metrics are commonly used. Accuracy is a basic metric that measures the proportion of correctly classified samples but may be misleading when dealing with imbalanced datasets. Precision, Recall, and F1-score provide a more detailed assessment: Precision measures the proportion of correctly predicted positive instances among all predicted positives, making it crucial in scenarios where false positives are costly (e.g., fraud detection). Recall (also called Sensitivity) evaluates the proportion of actual positive cases correctly identified, which is essential in applications like medical diagnosis, where missing a positive case can have severe consequences. The F1-score balances precision and recall, offering a more comprehensive evaluation when both false positives and false negatives need to be minimized. Additionally, AUC-ROC (Area Under the Receiver Operating Characteristic Curve) is a key metric for assessing a model’s ability to distinguish between classes across different threshold settings, particularly useful for comparing classifiers like SVM and XGBoost. Choosing the right evaluation metric depends on the specific application and the trade-offs between different types of classification errors.

\section{DAS Dataset}

\label{sec:3}
The dataset employed in this study was sourced from existing literature \cite{dataset}. It comprises strain rate measurements collected from May 16 to June 30, 2023, along the Albula River in Switzerland, specifically following the UNESCO World Heritage route of the Rhaetian Railway. The distributed acoustic sensing (DAS) setup spans a total fiber length of approximately 10 kilometers, with a measurement length of 8 meters and a channel spacing of 4 meters, resulting in a total of 512 channels. To optimize data processing, the DAS data were downsampled to a frequency of 20 Hz.
In  \cite{dataset}, the authors present a novel methodology that reimagines seismic event detection by transforming it into an image classification problem through an innovative data representation technique. Specifically, the distributed acoustic sensing (DAS) data are segmented into a series of 25.6-second intervals, each structured as a 512 × 512 pixel image. These generated images offer a comprehensive representation of strain rate time series, effectively capturing both temporal and spatial characteristics of the measured data.

By integrating geological landslide events documented through Doppler radar measurements, the data were systematically classified into four distinct noise categories: vehicle noise, random noise, slope failure, and narrow-band noise.

	The various categories contain sample data shown in Table \ref{tab.data}.
	
	\begin{table}[]
	
		\caption{Number of samples included in each category. }
			\label{tab.data}
		\centering
		\begin{tabular}{|l|l|}
				\hline
				
			          Category &	Counts\\ \hline	
			Vehicle noise	& 792 \\ \hline		
			Random noise& 123 \\ \hline			
			Slope failure&    384\\ \hline

			Narrow-band noise& 192  \\ \hline
         
		\end{tabular}
	\end{table}
	
\begin{figure}[htbp]
	\centering
	\includegraphics[width=8cm]{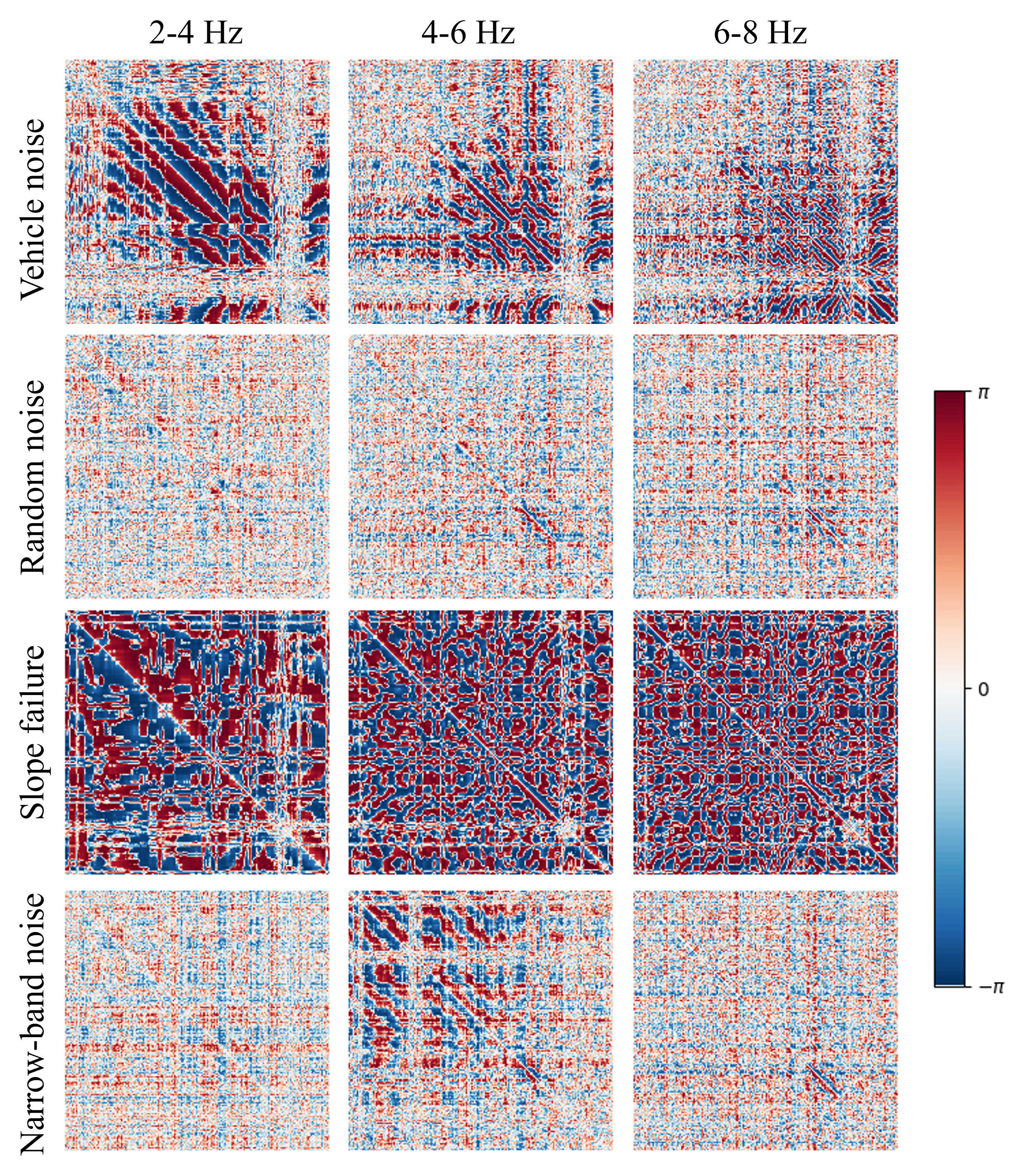}
	\caption{Various textures in the phase of CSDM.}
	\label{fig:example}
\end{figure}
\section{Experimental Results}
\label{sec:4}
\subsection{Data Preprocessing}
Cross-Spectral Density Matrices (CSDMs), denoted by $\phi_{CSDM}$, characterize the frequency-domain correlation between signals, which serve as a foundation for further feature extraction and are computed as follows:
\begin{equation}
	\phi_{CSDM}=<X(\omega),Y(\omega)>_{angle},
\end{equation}
where $X(\cdot)$ and $Y(\cdot)$ are the Fourier transforms of signal $x(t)$ and $y(t)$, respectively.
$<>_{angle}$ denote the complex phase and several samples are shown in Fig.~\ref{fig:example}.
We applied a diverse set of data augmentation techniques, including adjustments to brightness, contrast, hue, and the addition of Gaussian blur, to enhance the CSDM phase maps, resulting in a total of 22,365 samples. The labeled dataset was then partitioned into non-overlapping training, validation, and test sets following an 8:1:1 ratio.

Follow the reference \cite{dataset},
both labeled and unlabeled data were used to train the VAE \cite{vqvae}  for representation learning using large amounts of unlabeled data.
Each sample is compressed and represented as to 7 blocks, 512 channels, 2x2 space size.

\subsection{Benchmarking Methods}
We evaluated the recognition performance of multiple machine learning methods, including LR\cite{lr}, SVM\cite{svm}, KNN\cite{knn} and XGBoost\cite{xgboost} and MLP\cite{mlp}, ResNet\cite{resnet}, LSTM\cite{lstm}, Transformer\cite{transformer} and Mamba\cite{mamba2}.
To
mitigate the
 challenges posed by the curse 
of dimensionality inherent in classical methods,
we reduced the data dimensionality to 1000 when evaluating the performance of the classical approach.

\begin{figure*}[!t]
	\centering
	\subfigure[LR]{
		\includegraphics[width=3cm]{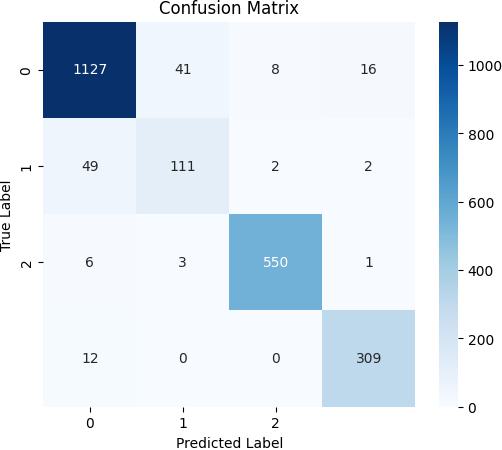}
	}
	\subfigure[SVM]{
		\includegraphics[width=3cm]{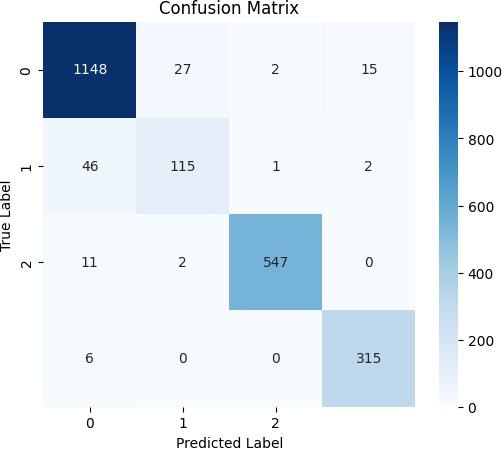}
	}
	\subfigure[KNN]{
		\includegraphics[width=3cm]{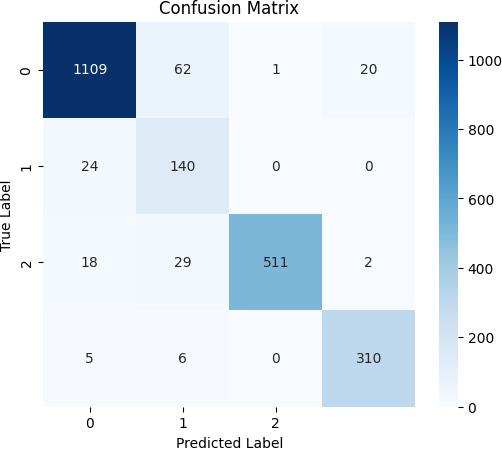}
	}
	\subfigure[XGBoost]{
		\includegraphics[width=3cm]{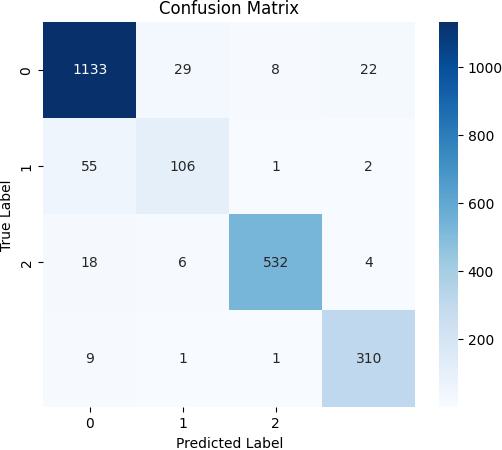}
	}
	\subfigure[MLP]{
		\includegraphics[width=3cm]{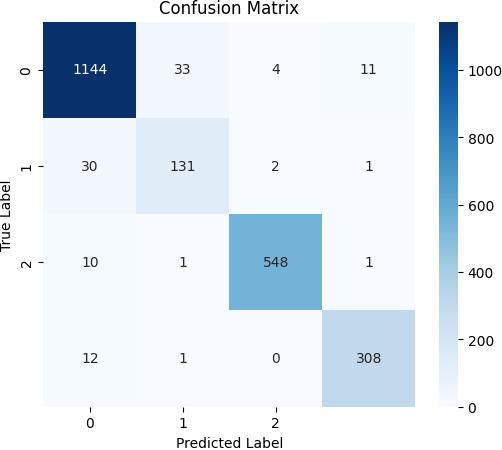}
	}
	\subfigure[CNN]{
		\includegraphics[width=3cm]{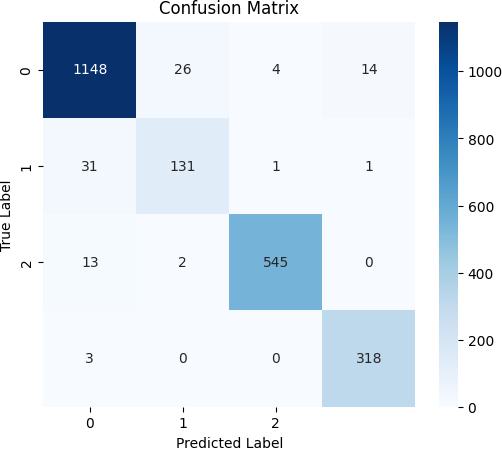}
	}
	\subfigure[LSTM]{
		\includegraphics[width=3cm]{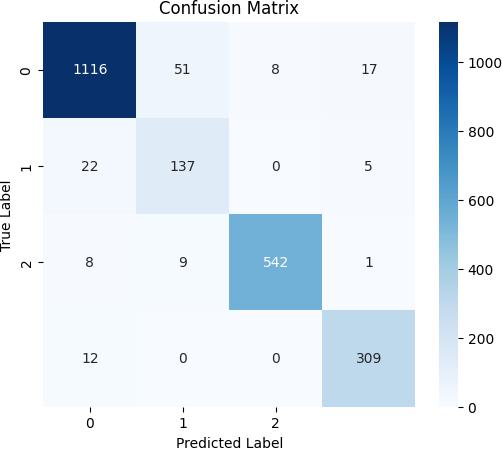}
	}
	\subfigure[Transformer]{
		\includegraphics[width=3cm]{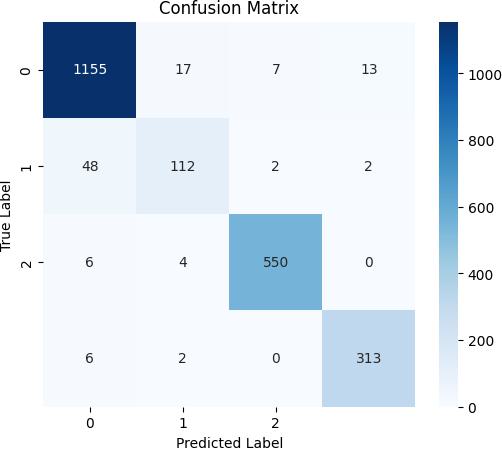}
	}
	\subfigure[Mamba]{
		\includegraphics[width=3cm]{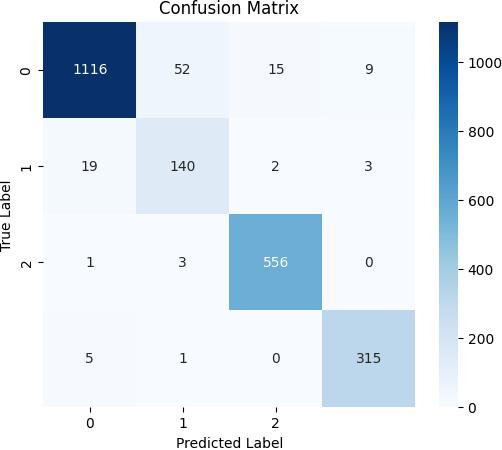}
	}
	\caption{Confusion matrix results.}
	\label{fig.cm}
\end{figure*}
\subsection{Recognition  Results}

Table \ref{tab.res} presents the classification performance of various machine learning and deep learning models in terms of accuracy and F1-score. Among traditional machine learning methods, SVM achieved the highest accuracy (0.950) and F1-score (0.913), outperforming LR, KNN, and XGBoost. Notably, deep learning models demonstrated superior performance, with ResNet achieving the best results overall, obtaining an accuracy of 0.958 and an F1-score of 0.932. Similarly, MLP and Transformer-based models exhibited competitive performance, with F1-scores of 0.924 and 0.914, respectively.

Comparing recurrent models, LSTM attained an accuracy of 0.941 and an F1-score of 0.908, indicating its ability to capture sequential dependencies. The Transformer model, however, slightly outperformed LSTM, suggesting its effectiveness in handling complex feature representations. Overall, deep learning approaches, particularly CNN-based architectures like ResNet, proved to be more effective than traditional machine learning models for this classification task.
The results of the confusion matrix are shown in Figure \ref{fig.cm}.

	\begin{table}[]
	
	\caption{Recognition  Results. }
	\label{tab.res}
	\centering
	\begin{tabular}{|l|l|l|}
		\hline
		
	Methods &Acc.&  F1-score            \\ \hline	
LR	 &0.937   &0.894 \\ \hline	
SVM	&  0.950   & 0.913\\ \hline	
KNN	&0.925&0.86 \\ \hline	
XGBoost	&0.930& 0.885\\ \hline	
MLP	&0.953& 0.924\\ \hline	
ResNet	&0.958& 0.932\\ \hline	
LSTM	&0.941&0.908 \\ \hline	
Transformer	&0.952& 0.914\\ \hline

	\end{tabular}
\end{table}

\section{Conclusion}
\label{sec:5}
In this study, we conducted a comprehensive benchmarking of various machine learning models for distributed acoustic sensing  classification, assessing their performance across multiple datasets and feature extraction techniques. Our findings reveal that deep learning-based approaches, particularly convolutional neural networks  and transformer-based architectures, consistently outperform traditional models such as logistic regression and support vector machines in terms of both accuracy and robustness.
Despite these advancements, several challenges remain. The need for computationally efficient models capable of real-time processing, the scarcity of labeled data in specific regions, and the sensitivity of models to environmental and seasonal variations continue to pose significant hurdles. Addressing these challenges will require further exploration of techniques such as transfer learning, domain adaptation, and multimodal fusion.
Future research directions may include developing strategies for handling class imbalance, leveraging federated learning for privacy-preserving DAS analysis, and enhancing model interpretability to improve deployment in real-world applications.
Our benchmarking study provides a foundation for the selection and optimization of machine learning models in DAS classification, paving the way for more robust, scalable, and efficient distributed acoustic sensing applications.

	\bibliographystyle{bib/IEEEtran.bst}
	\bibliography{bib/strings.bib}

\end{document}